\documentclass[aps,prl,twocolumn,showpacs]{revtex4}
\usepackage{graphicx}
\usepackage{amsfonts,amssymb,amsmath}
\usepackage{bm}
 
\begin{document}
 
\title{Weak ferromagnetism and magnetoelectric coupling in bismuth ferrite}

\date{\today}
 
\author{Claude Ederer}
\email{ederer@mrl.ucsb.edu}
\affiliation{Materials Research Laboratory and Materials Department,
  University of California, Santa Barbara, CA 93106, U.S.A.}
\author{Nicola A.~Spaldin}
\affiliation{Materials Research Laboratory and Materials Department,
  University of California, Santa Barbara, CA 93106, U.S.A.}

\begin{abstract}
We analyze the coupling between the ferroelectric and magnetic order
parameters in the magnetoelectric multiferroic BiFeO$_3$ using
density functional theory within the local spin density approximation
and the LSDA+U method. We show that weak ferromagnetism of the
Dzyaloshinskii-Moriya type occurs in this material, and we
analyze the coupling between the resulting magnetization and the
structural distortions. We explore the possibility of
electric-field-induced magnetization reversal and show that, although
it is unlikely to be realized in BiFeO$_3$, it is not in general
impossible. Finally we outline the conditions that must be fulfilled
to achieve switching of the magnetization using an electric field.

\end{abstract}

 
\maketitle

There has been increasing recent interest in {\it magnetoelectric
multiferroics}
\cite{Fiebig_Nature:2002,Wang/Neaton_Science:2003,Kimura:2003,Kimura_Nature:2003,Sudak:2004},
which are materials that show spontaneous magnetic and electric
ordering in the same phase. In addition to the fascinating physics
resulting from the independent existence of two or more ferroic order
parameters in one material \cite{Hill_review:2002}, the {\it coupling}
between magnetic and electric degrees of freedom gives rise to
additional phenomena. The linear and quadratic magnetoelectric (ME)
effects, in which a magnetization linear or quadratic in the applied
field strength is induced by an electric field (or an electric
polarization is induced by a magnetic field), are already well
established \cite{Sudak:2004}. Recently, more complex coupling
scenarios have been investigated. Examples are the coupling of the
antiferromagnetic and ferroelectric domains in hexagonal YMnO$_3$
\cite{Fiebig_Nature:2002} or the large magnetocapacitance near the
ferromagnetic Curie temperature in ferroelectric BiMnO$_3$
\cite{Kimura:2003}. Especially interesting are scenarios where the
direction of the magnetization or electric polarization can be
modified by an electric or magnetic field respectively. Such a
coupling would open up entirely new possibilities in data storage
technologies, such as ferroelectric memory elements that could be read
out nondestructively via the accompanying magnetization. Some progress
has been made in this direction. Recently, the small
(0.08$\mu$C/cm$^2$) electric polarization in perovskite TbMnO$_3$ was
rotated by 90$^\circ$ using a magnetic field at low temperatures
($\sim$10-20K) \cite{Kimura_Nature:2003}. Conversely, early work on
nickel-iodine boracite \cite{Ascher:1966} showed that, below
$\sim$60K, reversal of the spontaneous electric polarization rotates
the magnetization by 90$^\circ$ indicating that the axis of the
magnetization, but not its sense, can be controlled by an electric
field. In fact, it was believed \cite{Schmid:1999} that
electric-field-induced 180$^\circ$ switching of the magnetization
should be impossible, because a reversal of the magnetization
corresponds to the operation of time-inversion whereas the electric
field is invariant under this operation. In this work we show that
such behavior is not generally impossible by using multiferroic
bismuth ferrite, BiFeO$_3$, as a test case to analyze the coupling
between magnetism and ferroelectricity.

\begin{figure}[bthp]
\includegraphics*[width=0.3\textwidth]{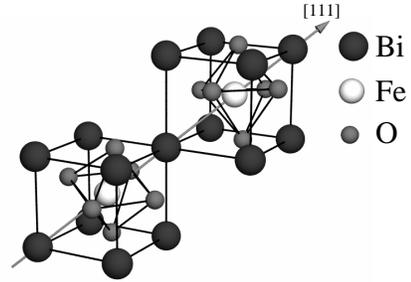}
\caption{Schematic view of the $R3c$ structure built up from two cubic
perovskite unit cells. The cations are displaced along the [111]
direction relative to the anions and the oxygen octahedra rotate with
alternating sense around the [111] axis.}
\label{fig:BFO}
\end{figure}

BiFeO$_3$ has long been known in its bulk form to be an
antiferromagnetic, ferroelectric multiferroic
\cite{Kiselev_BFO_AFM:1963,Teague_BFO_ferroelectric:1970}, with
antiferromagnetic N\'{e}el temperature $T_\text{N}\sim$643K, and
ferroelectric Curie temperature $T_\text{C}\sim$1103K. It has a
rhombohedrally distorted perovskite structure with space group $R3c$
\cite{Michel_BFO_structure:1969,Kubel/Schmid_BFO_structure:1990} (see
Fig.~\ref{fig:BFO}). The Fe magnetic moments are coupled
ferromagnetically within the pseudo-cubic (111) planes and
antiferromagnetically between adjacent planes (so-called G-type
antiferromagnetic order). If the magnetic moments are oriented
perpendicular to the [111] direction, the symmetry also permits a
canting of the antiferromagnetic sublattices resulting in a
macroscopic magnetization; so called weak ferromagnetism
\cite{Dzyaloshinskii:1957,Moriya:1960}. However, superimposed on the
antiferromagnetic ordering, there is a spiral spin structure in which
the antiferromagnetic axis rotates through the crystal with an
incommensurate long-wavelength period of $\sim$620\AA\
\cite{Sosnowska_BFO_spiral:1982}. This spiral spin structure leads to
a cancellation of the macroscopic magnetization and also inhibits the
observation of the linear ME effect
\cite{Popov_BFO_flop:1993}. However, a significant magnetization
($\sim$1$\mu_\text{B}$ per unit cell), as well as a strong ME
coupling, have been reported recently in high quality epitaxial thin
films \cite{Wang/Neaton_Science:2003}. This observation suggests that
the spiral spin structure is suppressed in thin films, perhaps due to
epitaxial constraints or enhanced anisotropy. Since these epitaxial
films also show large electric polarization ($\sim$50-60$\mu$C/cm$^2$)
they are promising candidate materials for ME device applications. In
this work we present results of first principles calculations of the
magnetic properties of BiFeO$_3$ and analyze the coupling between the
magnetic and ferroelectric properties in this material.

Our approach is based on density functional theory (DFT), see
e.g. \cite{Jones/Gunnarsson_DFTreview:1989}, and we use two different
implementations to cross-check our results; the projector-augmented
plane-wave (PAW) method \cite{Bloechl_PAW:1994} implemented in the
{\it Vienna Ab-initio Simulation Package} (VASP)
\cite{Kresse/Furthmueller_VASP1:1996,Kresse/Joubert_PAW:1999} and the
linear muffin-tin orbital method in the atomic sphere approximation
(LMTO-ASA) \cite{Andersen_LMTO:1975}, extended for the treatment of
non-collinear spin-configurations and spin-orbit coupling
\cite{Grotheer/Ederer/Faehnle_spinwaves:2001,Ederer_XMCD:2002}. These
two features, which are also implemented in the VASP code, are often
omitted in DFT calculations, but are essential for our investigations.
Except where explicitly stated, we use the crystal structure obtained
by optimizing the atomic positions within the experimentally observed
$R3c$ symmetry (see Fig.~\ref{fig:BFO}). Our calculated structural
parameters are identical (within the usual numerical accuracy) with
those given in \cite{Wang/Neaton_Science:2003} and agree well with the
experimentally observed structure
\cite{Kubel/Schmid_BFO_structure:1990}. To reproduce the situation in
the thin films we suppress the spiral spin structure in our
calculations. We also use two different treatments of the
exchange-correlation functional; the standard local spin-density
approximation (LSDA, see \cite{Jones/Gunnarsson_DFTreview:1989}) and
the LSDA+U method \cite{Anisimov:1997}. The LSDA+U method introduces
two parameters into the treatment of the Fe $d$ states, the Hubbard
parameter, $U$, and the exchange interaction, $J$. We use a value of
$J=1$eV and treat $U$ as a free parameter, varying it from 1eV to 7eV,
while keeping the structure fixed to that calculated within the
LSDA. For the PAW calculations we use potentials with 15 valence
electrons for Bi ($5d^{10}6s^26p^3$), 14 for Fe ($3p^63d^64s^2$), and
6 for O ($2s^22p^4$) and a plane wave cutoff of 400eV. For the
Brillouin-zone integrations we use a 5$\times$5$\times$5
Monkhorst-Pack {\bf k}-point mesh \cite{Monkhorst/Pack:1976} and the
tetrahedron method with Bl{\"o}chl correction
\cite{Bloechl_tetrahedron:1994} (both PAW and LMTO). These values
result in good convergence for all quantities under consideration.

Since the nature of weak ferromagnetism in BiFeO$_3$ is not well
established due to the presence of the spiral spin structure in the
bulk, we first investigate the occurrence and origin of weak
ferromagnetism, which is intimately connected with the symmetry of the
system \cite{Dzyaloshinskii:1957}. In BiFeO$_3$ it can only occur if
the sublattice magnetizations are oriented in the (111) plane so that
the symmetry is reduced to the magnetic space group $Bb$ or $Bb'$
which (apart from the primitive translations) contains only one glide
plane. In this case a canting of the magnetic sublattices does not
lead to a further reduction in symmetry and weak ferromagnetism can
occur. Therefore we first determine the preferred orientation of the
sublattice magnetizations. We do this by calculating the energy
difference between the arrangements with (i) magnetic moments aligned
parallel/antiparallel to the [111] direction and (ii) magnetic moments
oriented within the (111) plane. Both our methods result in an LSDA
energy difference of about 2meV, with a preferred orientation of the
magnetic moments within the (111) plane.  This arrangement is
compatible with the existence of weak ferromagnetism. Within the (111)
plane, orientations of the sublattice magnetizations parallel or
perpendicular to the glide plane are energetically equivalent. The
anisotropy energy is reduced to more realistic values within the
LSDA+U method, but the easy magnetization orientation is unchanged.

\begin{figure}[bthp]
\includegraphics*[width=0.35\textwidth]{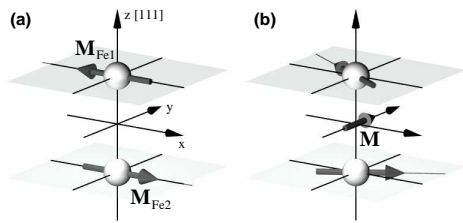}
\caption{(a): Starting configuration of our calculation. The magnetic
moments, ${\bf M}_\text{Fe1}$ and ${\bf M}_\text{Fe2}$, of the two
iron atoms in the unit cell are oriented antiferromagnetically and
collinearly in the (111) plane, allowing weak ferromagnetism by
symmetry. (b): Calculated magnetic structure including the spin-orbit
interaction: The two iron magnetic moments rotate in the (111) plane
so that there is a resulting spontaneous magnetization, ${\bf M}$.}
\label{fig:wfm}
\end{figure}

The anisotropy calculations show only that weak ferromagnetism is
symmetry-allowed, not that it will actually occur. Therefore we next
calculate the magnitude of the effect.  We initiate our calculations
to a homogeneous and collinear spin arrangement with the magnetic
moments oriented in the (111) plane (along either the $x$ or $y$-axis
in our coordinate system, see Fig.~\ref{fig:wfm}a), then let the
magnetic moments relax freely within the self-consistency cycle. The
magnetic moments then cant away from the collinear direction (while
remaining in the (111) plane) by an angle of about 1$^\circ$ (LSDA,
Fig.~\ref{fig:wfm}b). This leads to a small but measurable
magnetization of approximately $0.1 \mu_\text{B}$ per unit
cell. LSDA+U calculations give the same qualitative results but with
slightly smaller magnetizations. This value is smaller than that
reported in \cite{Wang/Neaton_Science:2003} but agrees well with more
recent measurements \cite{Mathur:PC}.

According to Dzyaloshinskii and Moriya (DM)
\cite{Dzyaloshinskii:1957,Moriya:1960} the canting of the magnetic
sublattices is caused by an antisymmetric spin coupling, the so-called
DM interaction, which is due to the combined action of exchange
interaction and spin-orbit coupling. Indeed, if we neglect the
spin-orbit interaction in our calculations the magnetic moments remain
collinear and there is no macroscopic magnetization. The DM
interaction has the form
\begin{equation}
\label{eq:DM-interaction}
E_\text{DM} = - \frac{1}{2} {\bf D} \cdot \left( {\bf M}_\text{Fe1}
\times {\bf M}_\text{Fe2} \right) = - {\bf D} \cdot \left( {\bf L}
\times {\bf M} \right) \quad ,
\end{equation}
where ${\bf D}$ is a coupling vector analogous to the Heisenberg
exchange constant $J$ in the usual symmetric exchange interaction. The
antiferromagnetic vector ${\bf L} = {\bf M}_\text{Fe1} - {\bf
M}_\text{Fe2}$ is defined as the difference of the two sublattice
magnetizations, and ${\bf M} = {\bf M}_\text{Fe1} + {\bf
M}_\text{Fe2}$ is the resulting magnetization. From the form of
$E_\text{DM}$ it is clear that, for constant ${\bf D}$ and fixed
orientation of ${\bf L}$, the canting of the magnetic sublattices
always occurs such that ${\bf D}$ (required by symmetry to be oriented
along the [111] axis), ${\bf L}$, and ${\bf M}$ build up a
right-handed system. Indeed, if we start our calculation with a spin
configuration in which the magnetic moments are canted in the “wrong”
direction (so that ${\bf D}$, ${\bf L}$, and ${\bf M}$ make up a
left-handed system) the moments relax back into the right-handed
configuration during the iteration process. Therefore, for a
particular orientation of ${\bf D}$ and ${\bf L}$, only one canting
direction lowers the energy relative to the collinear state.

Next we analyze the relationship between the weak ferromagnetism and
the structural distortions in BiFeO$_3$. As already mentioned, weak
ferromagnetism depends crucially on the symmetry of the system, which
in turn is determined by the structural distortions. The ferroelectric
$R3c$ structure of BiFeO$_3$ is reached from the ideal cubic
perovskite structure by freezing in two unstable normal modes: (i) the
polar displacements of all the anion and cation sublattices relative
to each other, which lead to the spontaneous electric polarization,
and (ii) an antiferrodistortive rotation of the oxygen octahedra
around the [111] direction with alternating sense of rotation along
the [111] axis (see Fig.~\ref{fig:BFO}). In terms of symmetry groups,
the polar displacements alone would reduce the symmetry of the ideal
perovskite structure ($Pm\bar{3}m$) to the rhombohedral space group
$R3m$, whereas the rotation of the oxygen octahedra alone would lead
to space group $R\bar{3}c$. The incorporation of both distortions
gives the actual space group of BiFeO$_3$, $R3c$. Weak ferromagnetism
is only allowed by symmetry in the space groups $R\bar{3}c$ and $R3c$,
suggesting that it is related to the oxygen rotations rather than to
the polar displacements along [111]. We have verified this by
performing calculations for structures containing only one of the two
distortions (while keeping the lattice vectors fixed to those of the
$R3c$ structure). These calculations confirm that the polar $R3m$
structure does not show weak ferromagnetism, whereas the
non-ferroelectric $R\bar{3}c$ does.

To fully understand the coupling between the structural distortions
and the magnetization, we next invert the sense of rotation of the
oxygen octahedra while keeping the polar distortions fixed. Again we
start from the magnetic configuration shown in Fig.~\ref{fig:wfm}a,
i.e. with the same orientation of ${\bf L}$ as in the previous
calculations. In this case the magnetization direction is reversed
from that of the original structure. Conversely, if we invert the
polar distortion while keeping the rotational sense of the oxygen
octahedra fixed, the magnetization direction is unchanged. This
clearly shows that the direction of the DM vector ${\bf D}$ is
determined by the sense of rotation of the oxygen octahedra
surrounding the corresponding magnetic ions, rather than by the
direction of the polarization as suggested in \cite{Fox/Scott:1977}.

\begin{table}
\begin{tabular}{c|c|c|c|c||c|c|c|c|}
& $1$ & $\bar{1}$ & $1'$ & $\bar{1}'$ & $m$ & $2\, (=\bar{1}m)$ & $m'$
& $2'$ \\ 
\hline
$i$ & 1 & 2 & 3 & 4 & 5 & 6 & 7 & 8 \\
\hline
${\bf P}$ & $\odot$ & $\otimes$ & $\odot$ & $\otimes$ & $\odot$ &
$\otimes$ & $\odot$ & $\otimes$
\\
${\bf D}$ & $\odot$ & $\otimes$ & $\odot$ & $\otimes$ & $\otimes$ &
$\odot$ & $\otimes$ & $\odot$
\\
${\bf M}$ & $\rightarrow$  & $\rightarrow$ & $\leftarrow$ &
$\leftarrow$ & $\leftarrow$ & $\leftarrow$ & $\rightarrow$ &
$\rightarrow$
\\
${\bf L}$ & $\downarrow$ & $\uparrow$ & $\uparrow$ & $\downarrow$ &
$\downarrow$ & $\uparrow$ & $\uparrow$ & $\downarrow$
\end{tabular}
\caption{All possible orientation states with parallel/antiparallel
  direction of ${\bf M}$. The first row gives the lost symmetry
  element that maps state 1 onto state $i$. $\bar{1}$ stands for
  space-inversion, a prime indicates time-inversion, $m$ is the mirror
  plane parallel to the [111] direction, and 2 is a twofold axis
  perpendicular to $m$. The states 5-8 are the antiphase domains not
  included in Aizu's scheme (see text). The directions of ${\bf P}$,
  ${\bf D}$, ${\bf M}$, and ${\bf L}$ projected on the (111) plane are
  indicated by arrows. $\odot$ ($\otimes$) indicates orientation along
  the positive (negative) [111] axis.}
\label{tab:domains}
\end{table}

Next we investigate the switching possibilities of the system from one
stable orientation state to another by the application of an electric
field.  A general scheme for the derivation of all possible
orientation states in ferroic materials was developed by Aizu
\cite{Aizu:1970}, who showed that the stable states can be constructed
by applying all symmetry elements that are lost during the ferroic
phase transition to an arbitrary orientation state of the ferroic
phase. A complication arises in the case of BiFeO$_3$ because the
symmetry change between the nonferroic prototype phase (nonmagnetic
$Pm\bar{3}m$) and the final multiferroic phase ($R3c$) is not purely
ferroic in nature. The unit cell doubling, caused by the oxygen
rotations, transforms the mirror plane parallel to the [111] direction
into a glide plane, and leads to the formation of antiphase domains
which are not described by Aizu's scheme.  To include the antiphase
domains one has to consider the full space group symmetry instead of
only the point group symmetry; for BiFeO$_3$ this analysis leads to a
total of 96 degenerate orientation states. Here we limit our
discussion to the 8 different orientation states with magnetization
parallel/antiparallel to a fixed axis. These represent 180$^\circ$
switching of the magnetization, and could be isolated in practice by
lifting the in-plane degeneracy of the easy magnetization axis using
epitaxial strain. The resulting orientation states are listed in
Table~\ref{tab:domains}, indicated by the different orientations of
${\bf M}$, ${\bf L}$, ${\bf D}$ and ${\bf P}$, where ${\bf P}$ is the
electric polarization and the ``sense'' of the oxygen rotations is
indicated by ${\bf D}$.

The application of a polarization-reversing electric field could in
principle drive the system from initial state 1 into any of the
degenerate states 2, 4, 6, or 8. Of course in reality the system will
prefer to change into the state separated from the initial state by
the lowest energy barrier. A reversal of ${\bf L}$ is unlikely since
it involves a $\sim$180$^\circ$ rotation of the sublattice
magnetizations, which is hindered by the magnetic anisotropy; in
contrast the reversal of ${\bf M}$ requires only a small reorientation
of the magnetic moments. This reduces the probable outcomes to either
state 4 (in which both ${\bf D}$ and ${\bf M}$ reverse) or state 8 (in
which ${\bf D}$ and ${\bf M}$ are unchanged). Of these, state 8, in
which the magnetic ordering is unchanged compared to state 1, is the
most likely, since the reversal of the oxygen rotations is
energetically costly, and is not required by the reversal of ${\bf
P}$. It is clear, however, that the earlier argument
\cite{Schmid:1999} that the mutual invariance under time- and
space-inversion of electric field and magnetization inhibits the
possibility of electric-field induced magnetization reversal, does not
hold. While it is certainly correct that it is not possible to drive
the system from state $1$ to its time-conjugate state $1'$ using an
electric field, the state corresponding to $\bar{1}'$ is also present
in all ferroelectric ferromagnets. If the energy barrier for the
transition to $\bar{1}'$ or to any other orientation state with
reversed ${\bf M}$ {\it and} ${\bf P}$ is lower than the energy
barrier to all other degenerate orientation states, then the
magnetization will be invertable by an electric field.

From the above discussion we can extract three conditions that must be
fulfilled to achieve electric-field-induced magnetization reversal in
such a rhombohedrally distorted multiferroic perovskite: (i) the
rotational and polar distortions must be coupled, (ii) the degeneracy
between states 1, 2, 3, 4 and 5, 6, 7, 8 must be lifted, i.e. parallel
and antiparallel orientations of ${\bf D}$ and ${\bf P}$ must be
inequivalent, and (iii) there should be only one easy magnetization
axis in the (111) plane. The latter condition can easily be achieved
by straining the material in an appropriate way. Fulfillment of the
other two conditions will take further exploration but we are not
aware of any general restriction that would make this impossible.

In summary, we have shown that BiFeO$_3$ exhibits weak ferromagnetism
of the DM type if the spiral spin structure is suppressed. We have
also shown that the DM vector is determined by the rotations of the
oxygen octahedra rather than by the ferroelectric
polarization. Finally we have discussed the possible magnetoelectric
switching scenarios in BiFeO$_3$, and formulated conditions that must
be met to realize electric-field-induced magnetization reversal.

\begin{acknowledgments}
The autors thank H.~Schmid and D.~Vanderbilt for valuable
discussions. This work was supported by the MRSEC program of the
National Science Foundation under Award No. DMR00-80034.
\end{acknowledgments}

\bibliography{paper.bib}
 
\end{document}